\documentclass[12pt]{article}
\usepackage[pdftex]{color}
\usepackage{amsmath,amsthm,amssymb}
\usepackage{wasysym}
\usepackage{graphicx}
\usepackage{color}%
\usepackage[doublespacing]{setspace}
\usepackage{lipsum}
\usepackage[lmargin=2.5cm,rmargin=2.5cm,tmargin=3cm,bmargin=3cm]{geometry}
\usepackage{multirow}
\usepackage{hyperref}
\hypersetup{hidelinks}
\usepackage{mathtools}
\usepackage{breqn}
\usepackage{threeparttable}
\usepackage[utf8]{inputenc}
\usepackage{tocloft}
\usepackage{cite}
\usepackage{enumitem}
\usepackage{dirtytalk}
\setlist[itemize]{leftmargin=*}

\begin{document}

\begin{center}
{\rm \bf \Large{Persson's Theory of Purely Normal Elastic Rough Surface Contact: A Tutorial Based on Stochastic Process Theory}
}
\end{center}

\begin{center}
{\bf Yang Xu$^a$\footnote{Corresponding author: yang.xu@hfut.edu.cn}, Xiaobao Li$^b$, Qi Chen$^a$, Yunong Zhou$^c$\footnote{Corresponding author: yunong.zhou@yzu.edu.cn}}\\
{
$^a$School of Mechanical Engineering, Hefei University of Technology, Hefei, 230009, China\\
$^b$School of Civil Engineering, Hefei University of Technology, Hefei, 230009, China  \\
$^c$Department of Civil Engineering, Yangzhou University,
Yangzhou, 225127, China
}
\end{center}

\begin{abstract}
Persson's theory of purely normal rough surface contact was developed two decades ago during the study of tire-road interaction, and gradually became one of the dominant approaches to study the solid-solid interaction between rough surfaces. Contrary to its popular applications in various cross-disciplinary fields, the fundamental study of Persson's theory of contact attracted little attention from the tribology and contact mechanics communities. As far as the authors know, many researchers struggle to understand the derivation of the theory. Few attempts have been made to clarify the oversimplified derivation provided by Persson (Persson, 2001). The present work provides a detailed tutorial on Persson's theory, which does not simply follow the commonly adopted derivation initiated by Persson. A new derivation is given based on stochastic process theory, assuming that the variation of the random contact pressure with respect to scale is a \emph{Markov process}. We revisit the essential assumptions utilized to derive the diffusion equation, boundary conditions, drift and diffusion coefficients, and closed-form results. This tutorial can   serve as a self-consistent introduction for solid mechanicians, tribologists, and postgraduate students who are not familiar with Persson's theory, or who struggle to understand it.
\end{abstract}

\section{Introduction}
Here we state the contact mechanics problem behind Persson's theory, relate preliminary knowledge of the stochastic features of   random contact pressure, and give a brief overview of Persson's theory.

\subsection{Problem Statement}
\begin{figure}[h!]
  \centering
  \includegraphics[width=8cm]{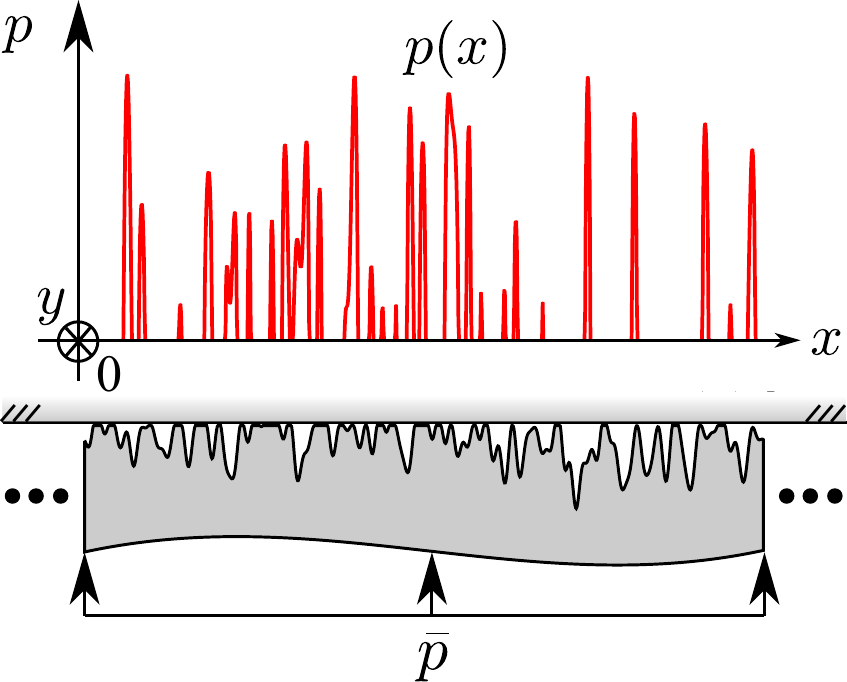}
  \caption{Graphical illustration of cross-sectional view of deformed interface when subjected to uniform compressive normal traction $\bar{p}$ and contact pressure distribution $p({\bf x})$ along line $y = 0$.}\label{fig:Fig_1}
\end{figure}

We study the contact problem  between a linear elastic half-space with a nominally flat rough surface and rigid flat. The rough surface topography is denoted by $h({\bf x})$, where ${\bf x} = (x, y) \in \mathbb{R}^2$. The topography is periodic along the in-plane $x$ and $y$ directions with respective periods $L_x$ and $L_y$. Therefore, the rough surface topography over one periodic in-plane domain of size $A_{\text{n}} = L_x \times L_y$ is representative of the entire topography. The rigid flat is fixed in space. The half-space is subjected to a uniform normal compressive traction $\bar{p}$ at the far end. The contact pressure distribution acting on the interface between the rigid flat and the half-space is $p({\bf x})$. Fig. \ref{fig:Fig_1} shows an example of a cross section of the contact interface. The interfacial shear stress, adhesion ($p < 0$), and third-body media (e.g., lubricant and wear particles) are not considered. 

\subsection{Preliminary knowledge}

Unlike other classic rough surface contact models \cite{archard1953contact,Greenwood66,Bush75,majumdar1990,ciavarella2000linear,jackson2006multi}, which heavily rely on asperity contact models from solid mechanics, Persson's theory is a stochastic process model in nature, which is relatively loosely related to solid mechanics. We relate some preliminary knowledge on commonly accepted stochastic features of random contact pressure $p({\bf x})$ to aid in understanding the derivation of the diffusion equation in section \ref{sec:FP_equ}.

The rough surface used in Persson's theory of contact is \emph{bandwidth-limited}, \emph{nominally flat}, \emph{isotropic}, and \emph{self-affine}. Let us define a parameter that describes the \emph{scale} at which the process is observed, $\zeta = q_{\text{s}}/q_{\text{l}}$, where $q_{\text{s}} = 2 \pi/\lambda_{\text{s}}$ ($\lambda_{\text{s}}$ is the shortest wavelength) and $q_{\text{l}} = 2 \pi/\lambda_{\text{l}}$ ($\lambda_{\text{l}}$ is the longest wavelength) are the respective upper and lower wavenumbers of the rough surface. A bandwidth-limited rough surface has a vanishing spectrum beyond these limits. The root mean square height of a nominally flat rough surface is significantly smaller than its lateral dimension (i.e., $L_x$ and $L_y$). An isotropic surface is referred to a special case where all profiles (along all in-plane directions) are statistically the same \cite{Manners06}. An mathematical definition of isotropy can be found in Eq. \eqref{eq:isotropy} in Appendix A. A self-affine rough surface implies that the amplitude to wavelength ratio cannot remain constant with the increasing wavenumber. A detailed discussion of the self-affinity of the rough surface is available in Appendix A. The spectral and statistical properties of this special rough surface group, which are extensively used in the rest of this tutorial, can be found in Appendix A. More detailed discussions of self-affine rough surfaces can be found  in  refs. \cite{Longuet57, Nayak71, Sayles78, Persson05, Jacobs17}. 

\begin{figure}[h!]
  \centering
  \includegraphics[width=14cm]{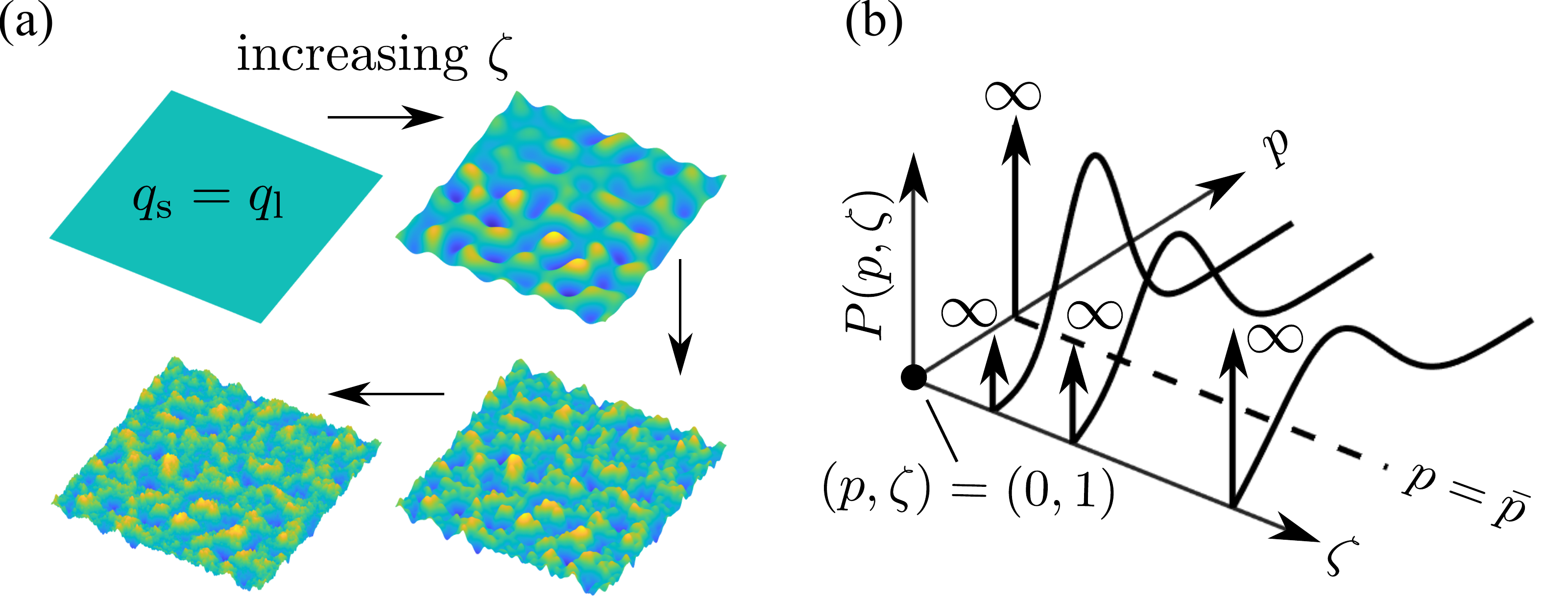}
  \caption{Evolution of: (a) nominally flat, isotropic, self-affine rough surface topography with respect to $\zeta$; (b)  PDF of contact pressure $P(p, \xi)$ with respect to $\xi$ as predicted by Persson's theory.}\label{fig:Fig_2}
\end{figure}

For a given scale $\zeta$, the corresponding contact pressure distribution and its probability density function (PDF) are denoted by $p({\bf x}, \zeta)$ and $P(p, \zeta)$, respectively. Under the non-adhesive assumption, the contact pressure can only take nonnegative values, so     $\forall \zeta \geq 1$, we have   conservation of   probability, 
\begin{equation}
\int_{0}^{\infty} P(p, \zeta) \text{d}p = 1.
\end{equation}
Let us define the contact and non-contact regions, where $p({\bf x}) > 0$ and $p({\bf x}) = 0$, respectively. As $\zeta \to 1$ (i.e., $q_{\text{s}} = q_{\text{l}}$), the entire power spectral density (PSD) of the rough surface is vanishing (see Eq. \eqref{E:PSD}). Hence, the rough surface becomes perfectly smooth, as seen in the top-left surface topography shown in Fig. \ref{fig:Fig_2}(a). Since the normal interaction between two flat surfaces results in   complete contact with  uniform contact pressure $p({\bf x}, \zeta = 1) = \bar{p}$, the corresponding $P(p, \zeta = 1)$ becomes a Dirac delta function (see the pulse with infinite magnitude on the $p$-axis in Fig. \ref{fig:Fig_2}(b)),
\begin{equation}\label{eq:initial_condition}
P(p, \zeta = 1) = \delta(p - \bar{p}).
\end{equation} 

As $\zeta$ increases, sinusoidal wavy components with higher wavenumbers ($q = 2 \pi/\lambda$, where $\lambda$ is the wavelength) and random phases are superposed onto the existing rough surface. The surface topography gradually becomes rougher as $\zeta$ increases (see Fig. \ref{fig:Fig_2}(a) for  graphical illustration). At the same time, $P(p, \zeta > 1)$ is \emph{broadened} from a singular pulse at $\zeta = 1$ to a \emph{bell-shaped} distribution with a larger pressure span over the real contact region ($p > 0$) (see Fig. \ref{fig:Fig_2}(b)). It is also expected that, as $\zeta$ increases, the complete contact achieved at $\zeta = 1$ cannot be maintained. Therefore, the probability of $p({\bf x}, \zeta > 1) > 0$, i.e., the area underneath $P(p > 0, \zeta > 1)$, is less than unity. The lost portion is completely ``squeezed" into the singular point at $p = 0$, forming a pulse of $P(p, \zeta)$, which represents the probability of zero pressure within the non-contact region (see pulses along the $\zeta$-axis in Fig. \ref{fig:Fig_2}(b)).

\begin{figure}[h!]
  \centering
  \includegraphics[width=8cm]{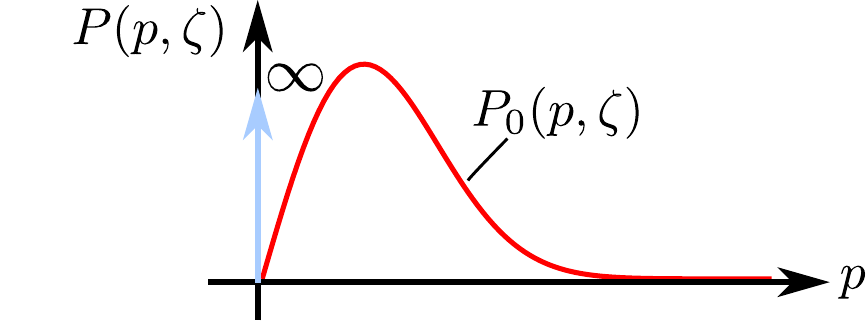}
  \caption{Piecewise representation of PDF $P(p, \zeta)$, including Dirac delta function $\delta(p)$ (purple arrow), and probability density function $P_0(p > 0,\zeta)$ (red solid line).}\label{fig:Fig_3}
\end{figure}

Based on the characteristics of $P(p, \zeta)$ given above, it may be regarded as a piecewise function with two parts \cite{Persson18, Xu22}: a Dirac delta function representing the probability of zero pressure at the non-contact region, and a PDF for compressive pressure within the contact region  (see Fig. \ref{fig:Fig_3} for  graphical illustration). Hence, $\forall p \in \mathbb{R}$,
\begin{equation}\label{eq:PDF_piecewise}
  P(p, \zeta) = \delta(p) \left[1 - A^*(\bar{p}, \zeta) \right] + \theta(p) P_0(p, \zeta),
\end{equation}
where $\delta(p)$ is the Dirac delta function; $A^*(\bar{p},\zeta)$ is the \emph{relative contact area} defined as the ratio of the real contact area $A_{\text{r}}$ to the nominal contact area $A_{\text{n}} = L_x \times L_y$, i.e.,
\begin{equation}
A^*(\bar{p}, \zeta) = \frac{A_{\text{r}}}{A_{\text{n}}} = \int_{0}^{\infty} P_0(p, \zeta) \text{d}p.
\end{equation}
The scale- and pressure-dependent $A^*(\bar{p}, \zeta)$ monotonically increases  with the increase of $\bar{p}$, and decreases with the increase of $\zeta$; $\theta(p)$ is the Heaviside function ($\theta(p > 0) = 1$ and $\theta(p < 0) = 0$)); $P_0(p,\zeta)$ is the PDF for compressive pressure only, which is continuous at $p = 0$, i.e., $P_0(p \to 0^+, \zeta) = P_0(p = 0, \zeta)$. Because of the Heaviside function in Eq. \eqref{eq:PDF_piecewise}, $P(p < 0, \zeta) = 0$, which is consistent with the non-adhesive assumption. Since the rough contact pair is subjected to  uniform compressive normal traction, the static load equilibrium in the normal direction is required:
\begin{equation}\label{eq:Load_equilibrium}
\bar{p} = \int_{0}^{\infty} p P_0(p, \zeta) \text{d}p.
\end{equation}

\subsection{A brief overview of Persson's theory}
Persson solved for $P_0(p, \zeta)$ in a diffusion equation, and gave the closed-form solution in terms of an infinite sum of sines \cite{Persson01}. In tribology, this probability-based contact mechanics approach is commonly referred to as Persson's theory of contact, and we demonstrate that this is also a stochastic process model. Persson's theory was originally used to solve for the purely normal and frictional contact between tire and road \cite{Persson01}, and has been extended to include more complex surface interaction laws (e.g., van der Waals force \cite{Persson01a, Persson02adhesion}, electrostatic force \cite{Persson18}, Lennard-Jones potential \cite{Joe17, Joe18}, friction \cite{Persson01, Scaraggi15}, and nonlinear deformation \cite{Persson01b, Papangelo21, Xu22}). This theory has been successively applied to   tribological problems, including electrical contact \cite{Persson22}, seals \cite{Bottiglione09, Persson16}, electroadhesion \cite{Persson18, Ciavarella20}, and contact electrification \cite{Persson20}. 

However, the fundamental study of Persson's theory has received little attention \cite{Persson01, Persson06, Manners06, Persson08, Muser08, prodanov2014contact, Dapp14, Xu22}, perhaps due to   difficulties   when reading Persson's original work \cite{Persson01}. Some difficulties can arise from: i) the oversimplified derivation of the diffusion equation with only five steps (Eqs. (B1)--(B5),   Appendix B, Ref. \cite{Persson01}); ii) unusual notation, concepts, and mathematical expressions (e.g., the characteristic function, PDF in terms of the Dirac delta function, and $\text{d}p$ placed in front of the integrand), which may be unfamiliar to   tribologists and solid mechanicians; and iii) indefinite integrals. These difficulties may prevent the full understanding of the fundamentals of Persson's theory of contact. We present a mathematically friendly (yet somewhat tedious) derivation of Persson's theory of purely normal elastic contact, based on stochastic process theory, so as to help non-physicists to   understand the fundamentals of Persson's theory. The current tutorial only discusses the evolution of $P_0(p, \zeta)$ with respect to $\zeta$, which is the main focus of Persson's classic work \cite{Persson01}. However, this is not the whole picture of Persson's theory. The mean gap vs. average contact pressure relation developed later by Persson and other researchers is not included \cite{PEI20052385, Persson07, Yang08, Akarapu11, prodanov2014contact, Papangelo17}. 

\section{Stochastic process model}\label{sec:FP_equ}
Persson's theory of contact belongs to stochastic process theory, which is the study of the evolution of the PDF of one or more random variables with state variables (e.g., time and scale)\cite{Gardiner04}. It has  wide application in statistical and quantum mechanics \cite{Risken89, Gardiner04, Gillespie92}, chemistry \cite{Kampen92}, and structural dynamics \cite{Li09}. A famous example of the stochastic process model can be found in Einstein's study of Brownian motion \cite{Einstein05}, where the diffusion equation of the probability of particle motion was derived. Like Einstein's derivation, implicit in Persson's original derivation of the diffusion equation (Appendix B, Ref. \cite{Persson01}) are  several important concepts of   stochastic process theory, e.g., the Chapman-Kolmogorov equation, Kramers-Moyal expansion, and   Fokker-Planck equation.

Further, we illustrate how to derive the Fokker-Planck equation from the Chapman-Kolmogorov equation for contact pressure. The diffusion equation in Persson's theory is a special case of the Fokker-Planck equation with zero drift coefficient. 

\subsection{Chapman-Kolmogorov equation}\label{subsec:CK_equation}

Consider two successive scales, $\zeta$ and $\zeta + \Delta\zeta$ (where $\Delta \zeta \geq 0$), at which the PDFs of the contact pressure over the entire contact interface are $P(p, \zeta)$ and $P(p, \zeta + \Delta \zeta)$, respectively. Assuming that the evolution of $P(p, \zeta)$ only depends on its most ``recent"\footnote{We make an analogy between the time-dependent stochastic process (e.g., Brownian motion), and the scale-dependent process  for rough surface contact.} scale, the transition from $P(p, \zeta)$ to $P(p, \zeta + \Delta \zeta)$ can be formulated as\footnote{The Chapman-Kolmogorov equation has also been used by Ciavarella et al. \cite{ciavarella2000linear} for solving a fractal rough profile in contact with a rigid flat.}
\begin{equation}\label{E:CK_forward}
    P(p, \zeta + \Delta \zeta) = \int_{0}^{\infty} P(p, \zeta + \Delta \zeta|p', \zeta) P(p', \zeta) \text{d} p',
\end{equation}
which is known as the \emph{Chapman-Kolmogorov equation} \cite{Gillespie92,Gardiner04}. The conditional probability $P(p, \zeta + \Delta \zeta| p', \zeta)\text{d}p$ denotes the probability of $p({\bf x}, \zeta + \Delta \zeta) \in [p, p + \text{d}p]$ given that $p({\bf x}, \zeta) = p'$. The conditional probability $P(p, \zeta + \Delta \zeta| p', \zeta)$ is commonly referred to as the \emph{transition probability density}. In the special case when $\Delta \zeta \to 0$, it follows that the transition probability density is the Dirac delta function, $\delta(p - p')$. The stochastic process in which the PDF evolution only relies on its most ``recent" history is referred to as the \emph{Markov process}. Hence, the fundamental assumption in Persson's theory is that \emph{the variation of the random contact pressure with respect to the scale is a Markov process}. 

\subsection{Transition probability density}
\begin{figure}[h!]
  \centering
  \includegraphics[width=10cm]{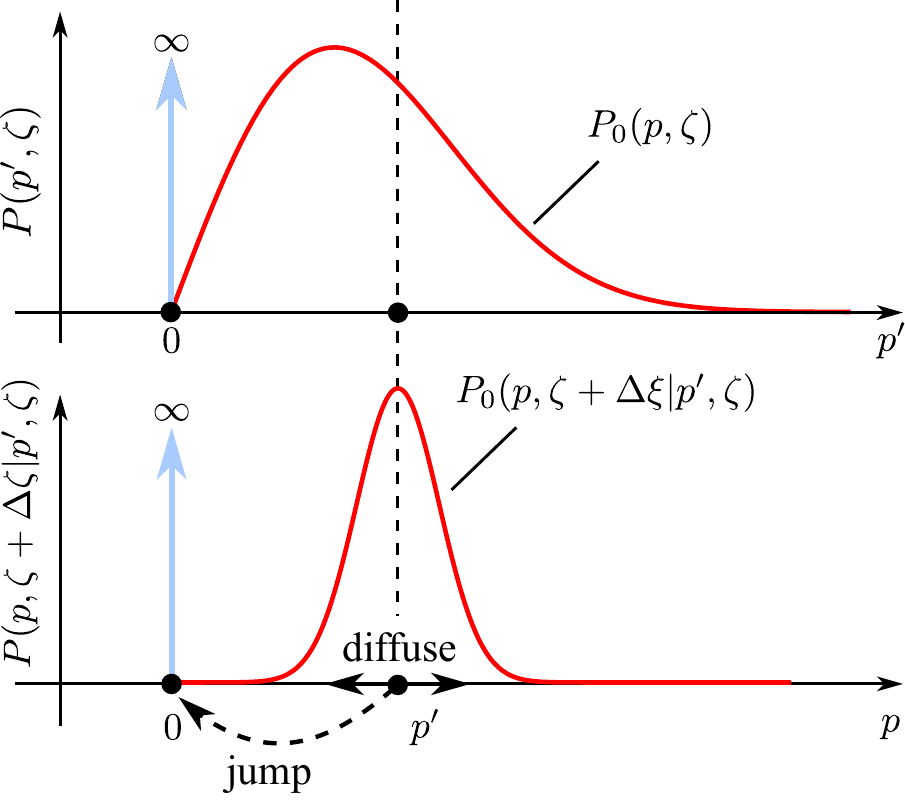}
  \caption{Graphical illustration of transition probability density $P(p, \zeta + \Delta \zeta | p', \zeta)$.}\label{fig:Fig_4}
\end{figure}

We further discuss the characteristics of $P(p, \zeta + \Delta \zeta|p', \zeta)$ with two cases: $p' > 0$ and $p' = 0$. In the former case, in which $p({\bf x}, \zeta) = p' > 0$, after an increase of $\zeta$ by $\Delta \zeta > 0$, the contact at $\mathbf{x}$ may be lost, meaning that the contact pressure has jumped from $p({\bf x}, \zeta) = p' > 0$ to $p({\bf x}, \zeta + \Delta \zeta) = p = 0$. It is also likely that this location remains in contact, but it can have any compressive pressure value with a certain probability. These two possibilities are associated with the \emph{jump} and \emph{diffusion} processes. The bottom part of Fig. \ref{fig:Fig_4} illustrates a typical transition probability density, which is composed of a jump term (blue pulse at $p = 0$) and diffusion term (red solid curve where $p > 0$). Inspired by the piecewise form of $P(p, \zeta)$ in Eq. \eqref{eq:PDF_piecewise}, the transition probability density associated with $p' > 0$ can be expressed similarly as
\begin{equation}\label{eq:Transition_Prob}
P(p, \zeta+\Delta \zeta |p' > 0, \zeta) = \delta(p) \left[1 - A_{\text{t}}^*(p', \zeta, \Delta \zeta) \right] + \theta(p) P_0(p, \zeta+\Delta \zeta |p', \zeta),
\end{equation}
where
\begin{equation}\label{eq:At}
A_{\text{t}}^*(p', \zeta, \Delta \zeta) = \int_0^{\infty} P_0(p, \zeta + \Delta \zeta | p', \zeta) \text{d} p.
\end{equation}
The ratio $A_{\text{t}}^*(p', \zeta, \Delta \zeta)$ represents the probability of an arbitrary location with $p({\bf x}, \zeta) = p'$ that still remains in contact after an incremental increase of $\zeta$. The following asymptotic behavior is extensively used in the later derivation:
\begin{equation}
\lim \limits_{\Delta \zeta \to 0} A_{\text{t}}^*(p', \zeta, \Delta \zeta) \to 1.
\end{equation} 
The transition probability is continuous at $p = 0$, i.e., $P_0(p \to 0^+, \zeta + \Delta \zeta|p', \zeta) = P_0(p = 0, \zeta + \Delta \zeta|p', \zeta)$. 

In the latter case, where $p'({\bf x}, \zeta) = p' = 0$, we assume that this location remains out-of-contact at any ``future" scale, i.e., 
\begin{equation}\label{eq:No_reentry}
P_0(p, \zeta + \Delta \zeta|p' = 0, \zeta) = \delta(p).
\end{equation} 
This is known as the \emph{no re-entry assumption}, which was implicit  in Persson's original derivation \cite{Persson01}, and was first found by Dapp et al. \cite{Dapp14}. This assumption states that an arbitrary location remains out-of-contact at ``future" scale once it is out-of-contact at the present scale. Under this assumption, the piecewise form in Eq. \eqref{eq:Transition_Prob} is also valid for $p' = 0$. Finally, it is easy to check that the transition probability density given in Eq. \eqref{eq:Transition_Prob} strictly satisfies   probability conservation, i.e., 
\begin{equation}
\int_0^{\infty} P(p, \zeta+\Delta \zeta |p' \geq 0, \zeta) \text{d} p = 1. 
\end{equation}

Replacing all PDFs in Eq. \eqref{E:CK_forward} with the piecewise forms given in Eqs. \eqref{eq:PDF_piecewise} and \eqref{eq:Transition_Prob}, two integral equations are finally obtained:
\begin{align}
A^*(\bar{p}, \zeta + \Delta \zeta) &= \int_0^{\infty} A_{\text{t}}^*(p', \zeta, \Delta \zeta) P_0(p', \zeta) \text{d} p', \label{eq:Area}\\
P_0(p, \zeta + \Delta \zeta ) &= \int_0^{\infty} P_0(p, \zeta + \Delta \zeta|p', \zeta) P_0(p', \zeta) \text{d} p', \label{eq:CK_P0}
\end{align}
which are of paramount importance in Persson's theory. Eq. \eqref{eq:Area} describes the evolution of the relative contact area due to an incremental increase of $\zeta$. For finite $p'$ and non-zero $\Delta \zeta$, $A_{\text{t}}^*(p', \zeta, \Delta \zeta) < 1$ in the R.H.S. of Eq. \eqref{eq:Area}. Thus, we can expect that $A^*(\bar{p}, \zeta + \Delta \zeta) < A^*(\bar{p}, \zeta)$, which implies that \emph{the relative contact area $A^*(\bar{p}, \zeta)$ monotonically decreases with the increase of $\zeta$ until it vanishes}. Eq. \eqref{eq:CK_P0} is the Chapman-Kolmogorov equation, which states that \emph{the variation of the random compressive contact pressure ($p > 0$) with the scale is also a Markov process}. Accordingly, we focus only on the stochastic process modeling of the random contact pressure within the contact region.

\subsection{Differential form of   Chapman-Kolmogorov equation}
The derivation of the differential form of the Chapman-Kolmogorov equation starts from an alternative form of Eq. \eqref{eq:CK_P0}. Applying a change of variable, $p' = p - \Delta p$, in the R.H.S. of Eq. \eqref{eq:CK_P0}, we   have
\begin{equation}\label{eq:CK_P0_alternative}
P_0(p, \zeta + \Delta \zeta ) = \int_{-\infty}^p P_0(p, \zeta + \Delta \zeta | p - \Delta p, \zeta) P_0(p - \Delta p, \zeta) \text{d} \Delta p.
\end{equation}
Let us define a new function,
\begin{equation}
    f(p) = P_0(p + \Delta p, \zeta + \Delta \zeta | p, \zeta) P_0(p, \zeta).
\end{equation}
Expanding $f(p - \Delta p)$ in  a Taylor series around $p$, we have
\begin{equation}\label{E:Taylor_expansion_forward}
    f(p - \Delta p) = f(p) + \sum_{n = 1}^{\infty} \frac{(-\Delta p)^n}{n!} \frac{\partial^n}{\partial p^n} f(p).
\end{equation}
Replacing the integrand in the R.H.S. of Eq. \eqref{eq:CK_P0_alternative} with Eq. \eqref{E:Taylor_expansion_forward},
\begin{align}\label{E:CK_forward_1}
    P_0(p, \zeta + \Delta \zeta) = & P_0(p, \zeta) \int_{-p}^{\infty} P_0(p + \Delta p, \zeta + \Delta \zeta|p, \zeta) \text{d} \Delta p ~ + \notag \\
    & \sum_{n=1}^{\infty} \frac{(-1)^n}{n!} \int_{-p}^{\infty} (\Delta p)^n \frac{\partial^n}{\partial p^n} \bigg[ P_0(p+\Delta p, \zeta + \Delta \zeta | p, \zeta) P_0(p, \zeta) \bigg] \text{d} \Delta p.
\end{align}
Dividing both sides of Eq. \eqref{E:CK_forward_1} by $\Delta \zeta$ and substituting $\int_{-p}^{\infty} P_0(p + \Delta p, \zeta + \Delta \zeta |p, \zeta) \text{d} \Delta p = A_{\text{t}}^*(p, \zeta, \Delta \zeta)$ (according to Eq. \eqref{eq:At}) into Eq. \eqref{E:CK_forward_1}, it can be simplified to
\begin{align}\label{E:CK_equation_2}
    &\frac{1}{\Delta \zeta} \left[ P_0(p, \zeta + \Delta \zeta) - A_{\text{t}}^*(p, \zeta, \Delta \zeta) P_0(p, \zeta) \right] = \notag \\
    &\frac{1}{\Delta \zeta} \sum_{n=1}^{\infty} \frac{(-1)^n}{n!} \int_{-p}^{\infty} (\Delta p)^n \frac{\partial^n}{\partial p^n} \left[ P_0(p+\Delta p, \zeta + \Delta \zeta | p, \zeta) P_0(p, \zeta) \right ] \text{d} \Delta p.
\end{align}
Now let us find the limiting form of Eq. \eqref{E:CK_equation_2} when $\Delta \zeta \to 0$. We first move the partial derivative operator outside Eq. \eqref{E:CK_equation_2}. Let $\Delta \zeta \to 0$, then $A_{\text{t}}^*(p, \zeta, \Delta \zeta) \to 1$ and $\lim \limits_{\Delta \zeta \to 0} P_0(0, \zeta + \Delta \zeta |p > 0, \zeta) = 0$. By mathematical induction, the integral in the R.H.S. of Eq. \eqref{E:CK_equation_2} with $\Delta \xi \to 0$ then can be rewritten as
\begin{align}\label{eq:partial_derivative}
&\int_{-p}^{\infty} (\Delta p)^n \frac{\partial^n}{\partial p^n} \left[ P_0(p+\Delta p, \zeta + \Delta \zeta | p, \zeta) P_0(p, \zeta) \right ] \text{d} \Delta p = \notag \\
&\frac{\partial^n}{\partial p^n} P_0(p, \zeta) \int_{-p}^{\infty} (\Delta p)^n P_0(p + \Delta p, \zeta + \Delta \zeta|p, \zeta) \text{d}\Delta p.
\end{align}
Finally, substituting Eq. \eqref{eq:partial_derivative} into Eq. \eqref{E:CK_equation_2} with $\Delta \zeta \to 0$, we obtain the \emph{Kramers-Moyal expansion},
\begin{equation}\label{E:Forward_KM_expansion}
\frac{\partial}{\partial \zeta} P_0(p, \zeta) = \sum_{n=1}^{\infty} \frac{(-1)^n}{n!} \frac{\partial^n}{\partial p^n} \left[B_n(p, \zeta) P_0(p, \zeta)\right],
\end{equation}
where $B_n(p, \zeta)$ is the rate of change of the $n^{\text{th}}$ moment of transition probability density with respect to $\zeta$,
\begin{equation}\label{E:Bn_forward_KM_expansion}
    B_n(p, \zeta) = \lim_{\Delta \zeta \to 0} \frac{1}{\Delta \zeta} \int_{-p}^{\infty} (\Delta p)^n P_0(p + \Delta p, \zeta + \Delta \zeta | p, \zeta) \text{d} \Delta p.
\end{equation}
Only keeping the partial derivative terms up to the second order, the Kramers-Moyal expansion reduces to the \emph{Fokker-Planck equation},
\begin{equation}\label{E:Fokker_Planck_Forward}
  \frac{\partial}{\partial \zeta} P_0(p, \zeta) = -\frac{\partial}{\partial p} \left[B_1(p, \zeta) P_0(p, \zeta)\right] + \frac{1}{2} \frac{\partial^2}{\partial p^2}\left[B_2(p, \zeta) P_0(p, \zeta)\right].
\end{equation}
The reason for only keeping the derivative terms up to the second order is explained in section \ref{subsec:summary} using Pawula's lemma \cite{Pawula67}). The scale- and pressure-dependent coefficients, $B_1(p, \zeta)$, and $B_2(p, \zeta)$, respectively, are referred to as the \emph{drift} and \emph{diffusion} coefficients. 

\section{Fokker-Planck equation}\label{sec:PDF_pressure}
The initial condition of the Fokker-Planck equation is given in Eq. \eqref{eq:initial_condition}. To obtain closed-form solutions, the drift and diffusion coefficients, and boundary conditions, should be known in advance.

\subsection{Drift and diffusion coefficients}\label{subsec:coefficients}
The Fokker-Planck equation with scale- and pressure-dependent drift and diffusion coefficients is difficult to solve analytically. To ease this complexity, we propose mean drift and diffusion coefficients,
\begin{equation}\label{eq:mean_Bn}
B_n(\zeta) = \int_{0}^{\infty} B_n(p, \zeta) P_0(p, \zeta) \text{d} p, ~~~ n = 1, 2.
\end{equation}
Replacing $B_n(p, \zeta)$ with $B_n(\zeta)$ in the Fokker-Planck equation (Eq. \eqref{E:Fokker_Planck_Forward}), the drift and diffusion coefficients can be taken outside the partial derivatives:
\begin{equation}\label{E:FP_Pressure_Partial_1}
\frac{\partial}{\partial \zeta} P_0(p, \zeta) = -B_1(\zeta) \frac{\partial}{\partial p} P_0(p, \zeta) + \frac{1}{2} B_2(\zeta) \frac{\partial^2}{\partial p^2} P_0(p, \zeta).
\end{equation}

Substituting Eq. \eqref{E:Bn_forward_KM_expansion} with $n = 1$ into Eq. \eqref{eq:mean_Bn}, we have (see Appendix C for a detailed derivation)
\begin{equation}
B_1(\zeta) = \frac{\text{d} \langle p \rangle}{\text{d} \zeta},
\end{equation}
where $\langle p \rangle(\zeta)$ is the mean contact pressure at scale $\zeta$. The mean drift coefficient $B_1(\zeta)$ represents the rate of change of the mean contact pressure with respect to the scale. According to the load equilibrium given in Eq. \eqref{eq:Load_equilibrium}, $\langle p \rangle(\zeta) = \bar{p}$. Hence, $B_1(\zeta) = 0$.

Substituting Eq. \eqref{E:Bn_forward_KM_expansion} with $n = 2$ into Eq. \eqref{eq:mean_Bn}, we have (see Appendix C for a detailed derivation)
\begin{equation}\label{eq:B2_xi}
B_2(\zeta) = \frac{\text{d} \langle p^2 \rangle}{\text{d} \zeta},
\end{equation}
where $\langle p^2 \rangle$ is the second moment of the contact pressure. Let us define the variance of contact pressure as $\text{Var}(p) = \langle (p - \bar{p})^2 \rangle$, where $\langle ~\rangle$ is the average symbol. Using the identity $\langle p^2 \rangle = \text{Var}(p) + \bar{p}^2$, the mean diffusion coefficient $B_2(\zeta)$ can be considered as the rate of change of the contact pressure variance with respect to the scale. Persson \cite{Persson01} assumed that $B_2(\zeta)$ can be approximated by its asymptotic form when the rough surface is in complete contact with the rigid flat, i.e., 
\begin{equation}\label{eq:B2_xi_1}
B_2(\zeta) = \frac{\text{d} \text{Var}(p)}{\text{d} \zeta} \approx \frac{\text{d} V}{\text{d} \zeta},
\end{equation}
where $V(\zeta)$ is the corresponding pressure variance (see Appendix B for detailed discussion). 

\subsection{Boundary conditions}

The boundary conditions used in the Fokker-Planck equation are 
\begin{equation}\label{E:BC_p0}
P_0(p = 0, \zeta) = 0, ~~~ P_0(p \to \infty, \zeta) = 0,
\end{equation}
where the latter one is naturally satisfied for a bounded PDF. The former one is known as the \emph{absorbing} boundary condition, which can be proved as follows \cite{Persson06}: applying $\int_0^{\infty} p \cdots \text{d}p$ to both sides of Eq. \eqref{E:FP_Pressure_Partial_1} with $B_1(\zeta) = 0$, we have $B_2(\zeta) P_0(p = 0, \zeta) = 0$. Since $B_2(\zeta) = \text{d}V/\text{d}\zeta$ is non-zero (see Eq. \eqref{eq:V_equ} for a complete expression of $V(\zeta)$), $P_0(p = 0, \xi)$ must vanish.

The absorbing boundary condition can   be explained more intuitively using Hertzian contact theory in a less rigorous way \cite{Dapp14}. The vanishing pressure is always at the edge of Hertzian contact area, and the corresponding pressure gradient is infinite. Assuming that the behaviour of the contact pressure at the edge of the contact region between the rough surface and the rigid flat is similar to that of Hertzian contact, it is easy to conclude that $P_0(p = 0, \zeta) = 0$. 

\subsection{Closed-form solutions}\label{subsec:Persson_original}
Substituting $B_1(\zeta) = 0$ and $B_2(\zeta) = \text{d} V/\text{d} \zeta$ into Eq. \eqref{E:FP_Pressure_Partial_1}, the well-known \emph{diffusion} equation in Persson's theory is obtained as
\begin{equation}\label{E:diffusion_equ}
\frac{\partial}{\partial \zeta} P_0(p, \zeta) = \frac{1}{2} \frac{\text{d} V}{\text{d} \zeta} \frac{\partial^2}{\partial p^2} P_0(p, \zeta).
\end{equation}
The initial and boundary conditions are
\begin{equation}\label{eq:DiffEqu_BCS}
P_0(p, \zeta = 1) = \delta(p - \bar{p}), ~~~ P_0(p = 0, \zeta) = 0, ~~~ P_0(p \to \infty, \zeta) = 0.
\end{equation}
The closed-form solution of $P_0(p, \zeta)$ was given first by Persson in terms of an infinite sum of sines (see Eq. (B16), Ref. \cite{Persson01}). Inspired by the work of Carslaw and Jaeger \cite{Carslaw59} on one-dimensional conduction of heat in solids, a much simpler form was given by Manners and Greenwood \cite{Manners06}, and independently by Yang et al. \cite{Yang06}.

\begin{figure}[h!]
  \centering
  \includegraphics[width=14cm]{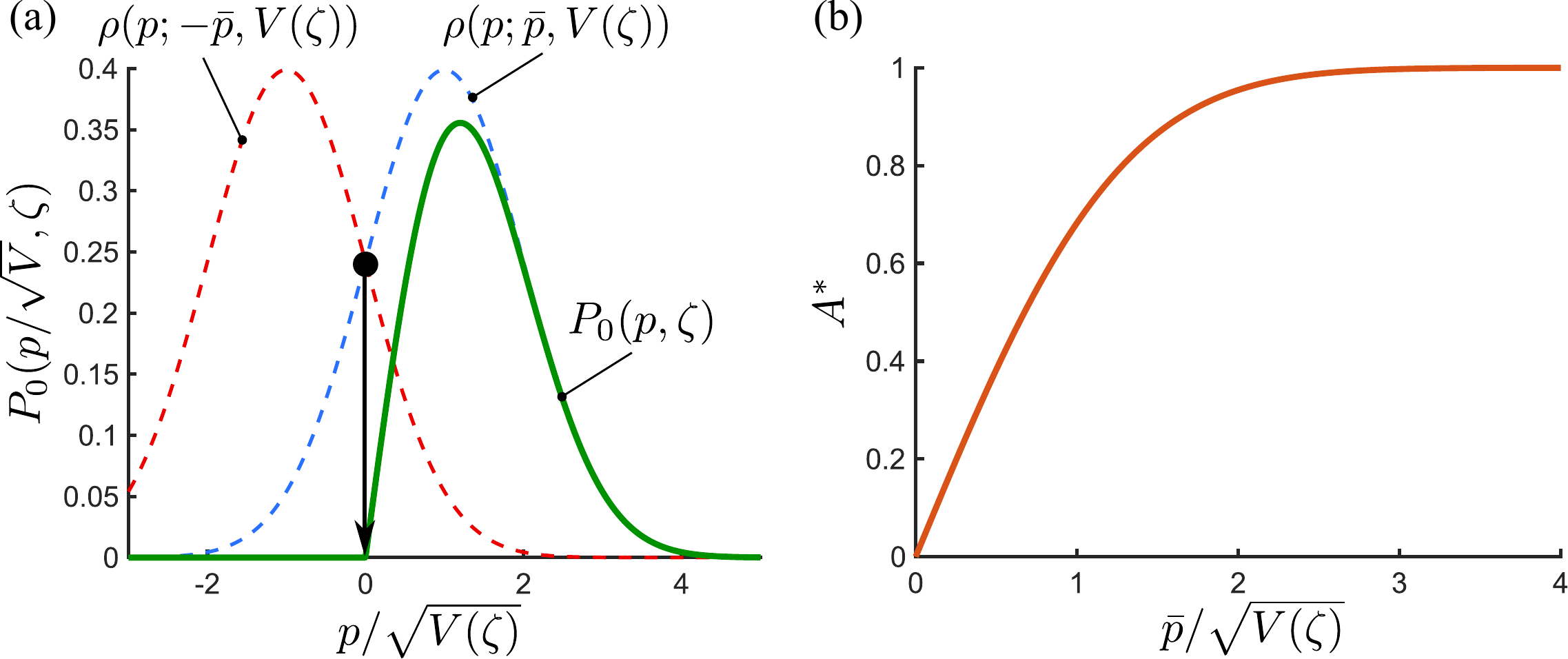}
  \caption{(a) Graphical illustration of   Gaussian distribution in Eq. \eqref{eq:Gauss_Distribution}, its mirror image about $p = 0$, and $P_0(p, \zeta)$ in Eq. \eqref{eq:P0_solution}; (b) Variation of $A^*(\bar{p}, \zeta)$ with $\bar{p}/\sqrt{V}$ predicted by Eq. \eqref{eq:Area_Persson}.}\label{fig:Fig_5}
\end{figure}

The diffusion equation has a special solution in the form of a Gaussian distribution \cite{Manners06}:
\begin{equation}\label{eq:Gauss_Distribution}
\rho(p; \bar{p}, V(\zeta)) = \frac{1}{\sqrt{2 \pi V(\zeta)}} \exp \left[ -\frac{(p - \bar{p})^2}{2 V(\zeta)} \right]. 
\end{equation}
The solution of the diffusion equation with initial and boundary conditions in Eqs. \eqref{eq:initial_condition} and \eqref{eq:DiffEqu_BCS} is composed of a Gaussian distribution, from which its mirror image about $p = 0$ is subtracted:
\begin{equation}\label{eq:P0_solution}
P_0(p, \zeta) = \rho(p; \bar{p}, V(\zeta)) - \rho(p; -\bar{p}, V(\zeta)).
\end{equation}
Fig. \ref{fig:Fig_5}(a)   illustrates how the absorbing boundary condition is strictly achieved using Eq. \eqref{eq:P0_solution}. 

Applying $\int_{0}^{\infty} \cdots \text{d} p$ to both sides of Eq. \eqref{eq:P0_solution}, the relative contact area is
\begin{equation}\label{eq:Area_Persson}
  A^*(\bar{p}, \zeta) = \text{erf} \left( \frac{\bar{p}}{\sqrt{2 V(\zeta)}}\right),
\end{equation}
where $\text{erf}(x) = \displaystyle{\frac{2}{\sqrt{\pi}} \int_0^x e^{-t^2} }\text{d}t$ is the error function. $A^*$ first linearly increases with $\bar{p}$, and then increases nonlinearly until it gradually converges to unity (see Fig. \ref{fig:Fig_5}(b)). Since $\displaystyle{\text{erf}(x)|_{x \to 0} \approx \frac{2}{\sqrt{\pi}} x}$, the asymptotic solution of Eq. \eqref{eq:Area_Persson} when $\bar{p} \to 0$ is \cite{Manners06}
\begin{equation}\label{eq:Asymptote_1}
A^*(\bar{p}, \zeta) \approx \sqrt{\frac{2}{\pi}} \frac{\bar{p}}{\sqrt{V(\zeta)}}.
\end{equation}
Using the identity $V = (E^*)^2 \langle |\nabla h|^2 \rangle/4$ given in Eq. \eqref{eq:VV_relation}, an equivalent form of Eq. \eqref{eq:Asymptote_1} is obtained as \cite{Yastrebov15}
\begin{equation}
A^*(\bar{p}, \zeta) \approx \sqrt{\frac{8}{\pi}} \frac{\bar{p}}{E^* \sqrt{\langle |\nabla h|^2 \rangle}}.
\end{equation}
Substituting Eqs. \eqref{eq:P0_solution} and \eqref{eq:Area_Persson} into Eq. \eqref{eq:PDF_piecewise}, 
\begin{equation}\label{eq:PDF_piecewise_explicit}
P(p, \zeta) = \delta(p) \left[1 - \text{erf} \left( \frac{\bar{p}}{\sqrt{2 V(\zeta)}}\right) \right] + \theta(p)\left[ \rho(p; \bar{p}, V(\zeta)) - \rho(p; -\bar{p}, V(\zeta)) \right].
\end{equation}

Persson's theory not only covers the evolution of $P(p, \xi)$ with respect to $\xi$, but also describes the mean gap $\bar{g}$ vs. $\bar{p}$ relation \cite{Persson07,Almqvist11,Afferrante18}. The derivation of $\bar{g}$ relies on the formulation of the elastic strain, $U_{\text{el}}$, stored in the deformed contact body. Using previous results of Persson's theory, $U_{\text{el}}$ can be obtained approximately \cite{Persson07, Persson08}. In the special case where the rough surface is completely flattened, $U_{\text{el}}$ is given in Eq. \eqref{eq:Uel_complete_1}. Assuming that the contact pressure distribution $p({\bf x})$ at partial contact ($A^* < 1$) is isotropic, i.e., the modulus of the Fourier transform is axisymmetric, $|\tilde{p}(q_x, q_y)| = |\tilde{p}\left(q = \sqrt{q_x^2 + q_y^2}\right)|$, $U_{\text{el}}$ can be obtained by simply replacing $|\tilde{p}_{\text{c}}(q)|^2$ with $|\tilde{p}(q)|^2$ in Eq. \eqref{eq:Uel_complete_1},
\begin{equation}\label{eq:Uel_form1}
U_{\text{el}} = \frac{(2 \pi)^3}{E^*} \int_{q_{\text{l}}}^{\zeta q_{\text{l}}} |\tilde{p}(q)|^2 \text{d} q.
\end{equation}
To obtain the analytical form of $|\tilde{p}(q)|^2$, we apply $\displaystyle{\int_0^{\infty} p^2 \cdots \text{d}p}$ to both sides of Eq. \eqref{E:diffusion_equ}:
\begin{equation}\label{eq:variance_p}
\frac{\partial }{\partial \zeta'} \text{Var}(p) = A^*(\bar{p}, \zeta') \frac{\text{d} V}{\text{d} \zeta'}.
\end{equation}
Defining $\zeta' = q/q_{\text{l}}$ and applying $\int_1^{\zeta} \cdots \text{d}\zeta'$ to both sides of Eq. \eqref{eq:variance_p}, we obtain  
\begin{equation}\label{eq:variance_p_form1}
\text{Var}(p) = \int_{q_{\text{l}}}^{\zeta q_{\text{l}}} A^*(\bar{p}, q/q_{\text{l}}) \frac{\text{d}}{\text{d} q} V(q/q_l) \text{d} q.
\end{equation}
The above equation utilizes the fact that $\text{Var}(p)|_{\zeta = 1} = 0$. Differentiating the integral form of $V$ given in Eq. \eqref{eq:V_equ} with respect to $q$,
\begin{equation}\label{eq:dVdxi}
\frac{\text{d}}{\text{d} q}V(q/q_{\text{l}}) = \frac{\pi}{2} (E^*)^2 q^3 C(q),
\end{equation}
where $C(q)$ is the PSD of the rough surface, see Eq. \eqref{E:PSD} for its piecewise expression. Inspired by Eq. \eqref{eq:variance_h_continuous}, Parseval's theorem leads to an alternative form of pressure variance:
\begin{equation}\label{eq:variance_p_form2}
\text{Var}(p) = \frac{(2 \pi)^3}{A_{\text{n}}} \int_{q_{\text{l}}}^{\zeta q_{\text{l}}} q |\tilde{p}(q)|^2 \text{d} q.
\end{equation}
Substituting Eq. \eqref{eq:dVdxi} into Eq. \eqref{eq:variance_p_form1} and equating the R.H.S. of Eqs. \eqref{eq:variance_p_form1} and \eqref{eq:variance_p_form2}, then the explicit form of $|\tilde{p}(q)|^2$ is \cite{Persson08}
\begin{equation}\label{eq:Pf_prime}
|\tilde{p}(q)|^2 = \left( \frac{E^*}{4 \pi} \right)^2 A_{\text{n}} q^2 C(q) A^*(\bar{p}, q/q_{\text{l}}).
\end{equation}
Replacing $|\tilde{p}(q)|^2$ in Eq. \eqref{eq:Uel_form1} with the R.H.S. of Eq. \eqref{eq:Pf_prime}, the elastic strain energy per nominal contact area is \cite{Persson08}
\begin{equation}\label{eq:Uel_original}
U_{\text{el}}(\bar{p}, \zeta) = \frac{\pi E^* A_{\text{n}}}{2}  \int_{q_{\text{l}}}^{\zeta q_{\text{l}}} q^2 C(q) A^*(\bar{p}, q/q_{\text{l}}) \text{d}q,
\end{equation}
where $A^*(\bar{p}, \zeta = q/q_{\text{l}})$ is determined by Eq. \eqref{eq:Area_Persson}. 

\section{Discussion}

\subsection{Summary of derivation}\label{subsec:summary}
Persson's theory of purely normal elastic rough surface contact is derived based on two fundamental assumptions and an approximation technique:

\begin{itemize}
\item 
The variation of random contact pressure ($p \geq 0$) with respect to the scale is a Markov process (see Eq. \eqref{E:CK_forward});

\item 
No re-entry assumption (see Eq. \eqref{eq:No_reentry});

\item 
Homogenization of $B_n(p, \zeta)$ over $p \geq 0$ (see Eq. \eqref{eq:mean_Bn}).
\end{itemize}
In Persson's original derivation (see Eq. (B1-B5) in Ref. \cite{Persson01}), and other similar works (e.g., Eqs. (23-28) in Ref. \cite{Manners06}), these aforementioned fundamental assumptions and approximation are not explicitly declared. In conventional stochastic process models (e.g., Brownian motion \cite{Einstein05}), the stochastic variable is spread across the entire real space. Due to the non-adhesive assumption, the stochastic variable $p$ is limited to non-negative values. This special range results in additional steps, where the partial derivative operator is moved outside the integral in Eq. \eqref{E:CK_equation_2}, but eventually lead to the same Kramers-Moyal expansion. The Fokker-Planck equation (Eq. \eqref{E:Fokker_Planck_Forward}) is obtained by truncating the Kramers-Moyal expansion (Eq. \eqref{E:Forward_KM_expansion}) at the second order derivative, but without providing a proof. To justify this truncation, we can refer to Pawula's lemma \cite{Pawula67}, which states: ``\emph{If $B_n$ as defined by Eq. \eqref{E:Bn_forward_KM_expansion}, exists for all $n$, and if $B_n = 0$ for some even $n$, then $B_n = 0$ for all $n \geq 3$}". Since it is impossible to guarantee that $B_{2n} \neq 0$ ($n > 1$), using the Kramers-Moyal expansion with finite terms ($n > 2$) is incorrect. Because solving the partial differential equation with an infinite number of derivative terms is impossible, the Fokker-Planck equation ($n = 2$) is the only viable option. However, we should keep in mind that the Fokker-Planck equation (Eq. \eqref{E:Fokker_Planck_Forward}) may not be a good approximation to the Chapman-Kolmogorov equation (Eq. \eqref{E:CK_forward}) \cite{Pawula67}.

\subsection{Fudge factors}
In section 3, the unknown diffusion coefficient is simply approximated by $B_2(\zeta) = \text{d}V/\text{d}\zeta$. This   only works when the rough surface is in \emph{nearly complete contact} with the rigid flat ($A^* \to 1$). For   medium and light load conditions, where the prediction of $A^*$ by Persson's theory is lower than that of numerical methods (e.g., the boundary element method \cite{Yastrebov15} and Green's function molecular dynamics \cite{prodanov2014contact}), this underestimation implies that the diffusion coefficient is overestimated by $B_2(\zeta) =\text{d}V/\text{d}\zeta$. Different empirical fudge factors $S(\bar{p}, \zeta)$ \cite{Yang08, Wang17}, which are curve-fitted from the numerical results, have been proposed to correct the diffusion coefficient as follows:
\begin{equation}\label{eq:new_B2}
B_2(\zeta) = S(\bar{p}, \zeta) \frac{\text{d} V}{\text{d} \zeta}.
\end{equation}
Yang and Presson \cite{Yang08} proposed the fudge factor
\begin{equation}
S(\bar{p}, \zeta) = \gamma + (1 - \gamma) \left[ A^*(\bar{p}, \zeta) \right]^2,
\end{equation}
with $\gamma \in [0.42, 0.48]$ \cite{Dapp14,Wang17,Almqvist11,Afferrante18}. An alternative form with a higher power of $A^*$ was given by Wang and M\"{u}ser \cite{Wang17, venugopalan2019plastic} as
\begin{equation}
S(\bar{p}, \zeta) = \gamma - \frac{2}{9} \left[ A^*(\bar{p}, \zeta) \right]^2 + \left(\frac{11}{9} - \gamma \right) \left[ A^*(\bar{p}, \zeta) \right]^4,
\end{equation}
where $\gamma = 5/9$. Replacing $\text{d}V/\text{d}\zeta$ in Eq. \eqref{E:diffusion_equ} with the R.H.S. of Eq. \eqref{eq:new_B2}, a nonlinear equation is obtained, which can  be solved iteratively for $A^*(\bar{p}, \zeta)$ as long as the history $A^*(\bar{p}, \xi' < \xi)$ is known in advance. 

\subsection{Compounded Chapman-Kolmogorov equation}
Besides its differential form (i.e., the diffusion equation), the Chapman-Kolmogorov equation can be iteratively applied along the scale direction to derive a \emph{compounded} Chapman-Kolmogorov equation \cite{Gillespie92},
\begin{equation}\label{E:Compounded_CK}
P(p, \zeta) = \int_{-\infty}^{\infty} \cdots \int_{-\infty}^{\infty} \prod \limits_{i=1}^{n} P(p_i, \zeta_i|p_{i-1}, \zeta_{i-1}) P(p_0, \zeta_0) \text{d} p_0 \cdots \text{d} p_{n-1},
\end{equation}
where $\boldsymbol{\zeta} = [\zeta_0, \zeta_1, \cdots, \zeta_n]^{\text{T}}$ is a vector of increasing scale with $\zeta_0 = 1$, $\zeta_n = \zeta$, and ${\bf p} = [p_0, p_1, \cdots, p_n]^{\text{T}}$ with $p_n = p$ is a vector of random variables so that $p({\bf x}, \zeta_i) = p_i$. The analytical form of the transition probability density was given by Dapp et al. \cite{Dapp14} as (see Appendix 2 in Ref. \cite{Wang17} for a detailed derivation)
\begin{align}\label{eq:transition_probability}
P(p, \zeta + \Delta \zeta|p', \zeta) \notag = ~&\delta(p) \left[1 - \text{erf} \left( \frac{p'}{\sqrt{2 \Delta V}} \right) \right] + \\
&\theta(p) \left[ \rho(p; p', \Delta V) - \rho(p; -p', \Delta V )\right],
\end{align}
where $\Delta V =  V(\zeta + \Delta \zeta) -  V(\zeta)$. Replacing $P(p_i, \zeta_i|p_{i-1}, \zeta_{i-1})$ and $P(p_0, \zeta_0)$ with the transition probability density in Eq. \eqref{eq:transition_probability} and initial condition in Eq. \eqref{eq:initial_condition}, respectively, the piecewise form of the PDF in Eq. \eqref{eq:PDF_piecewise_explicit} can be obtained through the mathematical induction. In a series of work by Joe et al. \cite{Joe17, Joe18, Joe22, JOE2023105397}, the compounded Chapman-Kolmogorov equation in terms of the PDF of the interfacial gap was developed to solve for the adhesive rough surface contact. 

\subsection{Zero contact area paradox}
In the end, we comment on the strange behavior of Persson's theory as $\zeta \to \infty$. According to Eq. (B.6), $V(\zeta) \sim \zeta^{-2H + 2}$. For $H < 1$, $V(\zeta)$ monotonically grows with the scale until the infinite value is reached at $\zeta \to \infty$. Hence, based on Eq. \eqref{eq:Area_Persson}, $A^*(\bar{p}, \zeta \to \infty) = 0$. Similar conclusion can also be drawn from Eq. \eqref{eq:Area}. Then, we can expect that the rough surface is completely separated from the rigid flat, while the far end of the rough surface is still subjected to the uniform normal traction of $\bar{p} > 0$. A similar paradox was found by Ciavarella et al. \cite{ciavarella2000linear}. This unrealistic asymptotic behavior may be solved using a finite scale, $\zeta$. However, there is a lack of methodology to guide the selection of $\zeta$. Fundamentally, this paradox may be due to the oversimplified surface interaction law (i.e., non-adhesive contact) and surface deformation model (i.e., linear elastic contact). Persson \cite{Persson01b} showed that the real contact area converges to a finite value as $\zeta \to \infty$ when the linear elastic deformation model is replaced by a truncated elastoplastic model with  constant hardness. Non-vanishing results at $\zeta \to \infty$ were also reported in Refs. \cite{Persson02adhesion, Joe17} if adhesion is included in the surface interaction law. 

\section{Conclusion}
Based on stochastic process theory, here we revisited Persson's theory of purely normal elastic contact. The random contact pressure variation with respect to the scale was assumed to follow a Markov process. We have derived the integral form of the Chapman-Kolmogorov equation for the PDF of compressive contact pressure within the contact area, assuming no re-entry. Using the homogenization of the drift and diffusion coefficients over the compressive pressure range, the diffusion equation of Persson's theory was obtained. The conventional derivation of the absorbing boundary condition and diffusion coefficient is provided. We revisited the detailed derivation of closed-form solutions for the PDF $P_0(p, \xi)$, relative contact area $A^*(\bar{p}, \xi)$, and elastic strain energy $U_{\text{el}}(\bar{p}, \xi)$. Several important topics are discussed, including fundamental assumptions, Pawula's lemma, fudge factors, the compounded Chapman-Kolmogorov equation, and the paradox of zero contact area at an infinite scale. To the best of our knowledge, this was the first comprehensive study of Persson's theory using the perspective of stochastic process theory. This tutorial can serve as a self-consistent introduction for solid mechanicians, tribologists, and postgraduate students who are not familiar with Persson's theory, or who struggle to understand it. We anticipate this tutorial will draw significant interest from the solid mechanics and tribology communities regarding the application of stochastic process theory in the field of rough surface contact.

\section*{Acknowledgements}
This work was supported by the National Natural Science Foundation of China (No. 52105179), Fundamental Research Funds for the Central Universities 
(No. JZ2023HGTB0252), Natural Science Foundation of Jiangsu Province (No. BK20220555), Natural Science Foundation of Anhui Province (No. 2208085MA17), and Anhui Provincial Key Research and Development Program (No. JZ2022AKKG0093). Authors would like to thank Prof. Andreas Almqvist (Luleå University of Technology, Sweden). Without his encouragement and careful editing of the manuscript, authors could not have produced this tutorial. YX would like to thank Dr. Junki Joe (National Renewable Energy Laboratory, USA) and Mr. Longan Zhu (Hefei University of Technology, China) for their helpful comments and constructive suggestions.

\begin{spacing}{1} 
	\bibliographystyle{asmejour}
	\bibliography{ref}
\end{spacing}

\appendix
\setcounter{figure}{0}
\setcounter{equation}{0}
\renewcommand{\thefigure}{A.\arabic{figure}}
\renewcommand{\theequation}{A.\arabic{equation}}
\renewcommand{\thetable}{A.\arabic{table}}

\section*{Appendix A. Nominally flat, isotropic, self-affine, rough surfaces}

Consider a periodic nominally flat rough surface $h(x, y)$ with periods $L_x$ and $L_y$ in the $x$ and $y$ directions, respectively. The projected area of the rough surface in one period on the $xy$ plane is $A_{\text{n}} = L_x \times L_y$. The inverse Fourier transform and Fourier transform for the periodic function are given respectively as 
\begin{align}
h(x, y) = &\frac{(2 \pi)^2}{A_{\text{n}}} \sum_{m = -\infty}^{\infty} \sum_{n = -\infty}^{\infty} \tilde{h}(m, n) \text{e}^{i 2 \pi \left( \frac{m}{L_x}x + \frac{n}{L_y}y \right)}, \label{eq:FT_inverse}\\
\tilde{h}(m, n) = &\frac{1}{(2 \pi)^2} \int_{-L_x/2}^{L_x/2} \int_{-L_y/2}^{L_y/2} h(x, y) \text{e}^{-i2 \pi \left( \frac{m}{L_x} x + \frac{n}{L_y} y \right)} \text{d} x \text{d} y, \label{eq:FT_forward}
\end{align}
where $\tilde{h}(m, n)$ is the discrete amplitude spectrum associated with the sinusoidal wavy component with \emph{angular wavevector} ${\bf q} = [q_x = 2 \pi m/L_x, q_y = 2 \pi n/L_y]$. When $L_x$ and $L_y$ are infinitely large \cite{prodanov2014contact, Jacobs17, Wang17}, the Fourier transform pair in Eqs. \eqref{eq:FT_inverse} and \eqref{eq:FT_forward} become
\begin{align}
h(x, y) = &\int_{-\infty}^{\infty} \int_{-\infty}^{\infty} \tilde{h}(q_x, q_y) \text{e}^{i \left( q_x x + q_y y \right)} \text{d} q_x \text{d} q_y, \label{eq:FT_inverse_infty}\\
\tilde{h}(q_x, q_y) = &\frac{1}{(2 \pi)^2} \int_{-\infty}^{\infty} \int_{-\infty}^{\infty} h(x, y) \text{e}^{-i \left( q_x x + q_y y \right)} \text{d} x \text{d} y. \label{eq:FT_infty}
\end{align}
The Fourier transform pair given in Eqs. \eqref{eq:FT_inverse_infty} and \eqref{eq:FT_infty} is denoted by $h(x, y) \Leftrightarrow \tilde{h}(q_x, q_y)$. 

Let us define the Power Spectral Density (PSD) of $h(x, y)$ as \cite{Jacobs17, Persson08}
\begin{equation}\label{eq:PSD_def}
S[h](q_x, q_y) = C(q_x, q_y) = \displaystyle{\frac{\left(2\pi\right)^2}{A_{\text{n}}}}|\tilde{h}(q_x, q_y)|^2.
\end{equation}
The unit of PSD is [mm$^4$]. For an \emph{isotropic} rough surface, a continuous PSD satisfies
\begin{equation}\label{eq:isotropy}
C(q_x, q_y) = C(q) = \frac{\left(2\pi\right)^2}{A_{\text{n}}} |\tilde{h}(q)|^2,
\end{equation}
where $q = \sqrt{q_x^2 + q_y^2}$.
\begin{figure}[h!]
  \centering
  \includegraphics[width=8cm]{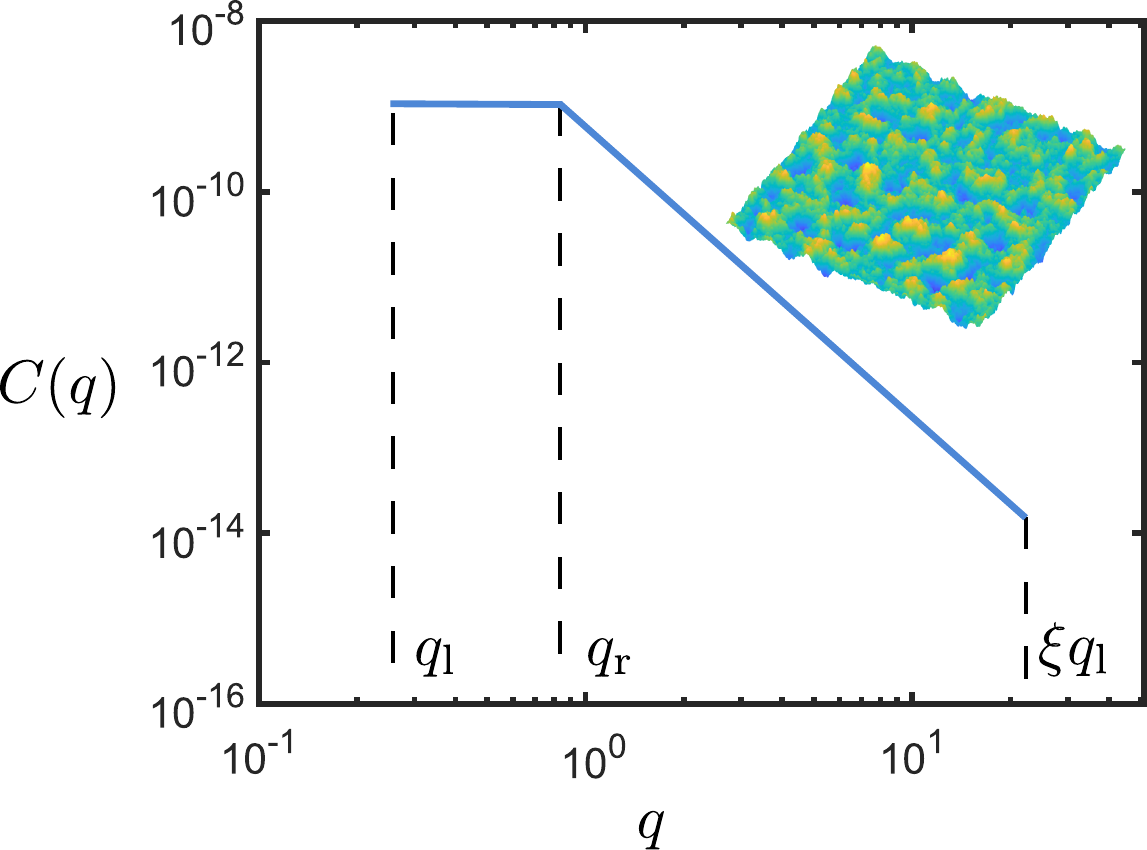}
  \caption{Schematic of PSD $C(q)$ of a bandwidth-limited isotropic rough surface.}\label{fig:Fig_A1}
\end{figure}

Commonly, the bandwidth-limited PSD of a nominally flat, self-affine, isotropic rough surface is expressed as
\begin{equation}\label{E:PSD}
  C(q) =
  \begin{cases}
  C_0 q_{\text{r}}^{-2 (1 + H)} ~~~ q \in [q_{\text{l}}, q_{\text{r}}), \\
  C_0 q^{-2 (1 + H)} ~~~ q \in [q_{\text{r}}, \zeta q_{\text{l}}], \\
  0 ~~~ \text{elsewhere},
  \end{cases}
\end{equation}
where $q_{\text{l}}$ and $\zeta q_{\text{l}}$ are the lower and upper cutoff wavenumbers, respectively. A typical PSD is plotted on the logarithmic scale in Fig. \ref{fig:Fig_A1}. The PSD within the lower wavenumber range remains constant, followed by a monotonic drop when $q > q_{\text{r}}$ (the roll-off wavenumber). $C_0$ is a proportionality constant. $H \in [0, 1]$ is the Hurst exponent. Assuming $\xi q_{\rm{l}} \gg q_{\rm{r}}, q_{\rm{l}}$, then if $H = 0$, the amplitude-to-wavelength ratio, $q |\tilde{h}(q)| = \displaystyle{\frac{q}{2 \pi}} \sqrt{A_{\text{n}} C(q)}$, remains constant $\forall q \in [q_{\rm{r}}, \xi q_{\rm{l}}]$; If $H > 0$, this ratio drops as $q$ increases. The rough surface with PSD in Eq. \eqref{E:PSD} is \emph{self-affine} if $H > 0$, and \emph{self-similar} if $H = 0$.

The definition of PSD given in Eq. \eqref{eq:PSD_def} results in the Fourier transform pair between the auto-correlation function and PSD, i.e., $\text{Corr}[h, h](x, y) \Leftrightarrow C(q_x, q_y)$, where
\begin{equation}
\text{Corr}[h, h](x, y) = \frac{1}{A_{\text{n}}} \int_{-L_x/2}^{L_x/2} \int_{-L_y/2}^{L_y/2} h(x' + x, y' + y) h(x', y') \text{d} x' \text{d} y'.
\end{equation}
The variance of the rough surface height can be denoted by $\langle h^2 \rangle = \text{Corr}[h, h](0, 0)$. Based on the Parseval's theorem, this scale-dependent variance can be written as
\begin{equation}\label{eq:variance_h_continuous}
\langle h^2 \rangle = \frac{(2 \pi)^3}{A_{\text{n}}} \int_{q_{\text{l}}}^{\zeta q_{\text{l}}} q |\tilde{h}(q)|^2 \text{d} q = 2 \pi \int_{q_{\text{l}}}^{\zeta q_{\text{l}}} q C(q) \text{d} q.
\end{equation}
Substituting Eq. \eqref{E:PSD} into Eq. \eqref{eq:variance_h_continuous}, we have the closed-form solution of $\langle h^2 \rangle$ \cite{Yastrebov15},
\begin{align}
\langle h^2 \rangle &=
\begin{cases}
\pi C_0 q_{\text{r}}^{-2 H - 2} q_{\text{l}}^2 \left( \zeta^2  -  1 \right) ~~~ \zeta q_{\text{l}} < q_{\text{r}},\\
\pi C_0 \left[ q_{\text{r}}^{-2H - 2} \left(q_{\text{r}}^2 - q_{\text{l}}^2 \right) + \displaystyle{\frac{1}{H}} \left( q_{\text{r}}^{-2H} - \zeta^{-2H} q_{\text{l}}^{-2H} \right) \right]  ~~~ \zeta q_{\text{l}} \geq q_{\text{r}}. \\
\end{cases} \label{eq:Vh_closed_form}
\end{align}
According to Eq. \eqref{eq:FT_inverse_infty}, we have $\partial h/\partial x + \partial h/\partial y \Leftrightarrow i (q_x + q_y) \tilde{h}(q_x, q_y)$. Based on the definition of PSD in Eq. \eqref{eq:PSD_def}, $S\left[\partial h/\partial x + \partial h/\partial y\right] = (q_x + q_y)^2 C(q_x, q_y)$. Using Parseval's theorem, we can have
\begin{equation}\label{eq:rms_slope_1}
\bigg \langle \left(\frac{\partial h}{\partial x} + \frac{\partial h}{\partial y} \right)^2 \bigg \rangle= \iint (q_x + q_y)^2 C(q_x, q_y) \text{d} q_x \text{d} q_y.
\end{equation}
Let $|\nabla h|^2 = (\partial h/\partial x)^2 + (\partial h/\partial y)^2$ be the modulus of surface gradient vector. Since $\partial h/\partial x$ and $\partial h/\partial y$ are uncorrelated \cite{Manners06} (i.e.,  $\langle (\partial h/\partial x) (\partial h/\partial y) \rangle = 0$), $\langle (\partial h/\partial x + \partial h/\partial y)^2 \rangle = \langle |\nabla h|^2 \rangle$. Because $C(q_x, q_y) = C(q)$ is an even function, we can further simplify Eq. \eqref{eq:rms_slope_1} to \cite{Afferrante18}
\begin{equation}\label{eq:V_gradient}
\langle |\nabla h|^2 \rangle = 2 \pi \int_{q_{\text{l}}}^{\zeta q_{\text{l}}} q^3 C(q) \text{d} q,
\end{equation}
where $\sqrt{\langle |\nabla h|^2 \rangle}$ is commonly known as the root mean square surface slope.

\setcounter{figure}{0}
\setcounter{equation}{0}
\renewcommand{\thefigure}{B.\arabic{figure}}
\renewcommand{\theequation}{B.\arabic{equation}}
\renewcommand{\thetable}{B.\arabic{table}}

\section*{Appendix B. Complete Contact}
When an elastic half-space with a nominally flat, isotropic, self-affine rough boundary is completely flattened by a rigid flat, the normal surface displacement is the same as the rough surface height $h(x, y)$ if the rigid body displacement is ignored. The resultant contact pressure distribution is represented by $p(x, y) = \bar{p} + p_{\text{c}}(x, y)$ with $\langle p_{\text{c}} \rangle = 0$. The relation between $h(x, y)$ and $p_{\text{c}}(x, y)$ can be characterized using the following convolution \cite{johnson1987contact}:
\begin{equation}\label{E:Boussinesq}
  h(x, y) = \frac{1}{\pi E^*} \int_{-\infty}^{\infty} \int_{-\infty}^{\infty} \frac{1}{\sqrt{(x - x')^2 + (y - y')^2}} p_{\text{c}}(x', y') \text{d} x' \text{d} y'.
\end{equation}
The plane strain modulus $E^* = E/(1 - \nu^2)$. Young's modulus and Poisson's ratio are denoted by $E$ and $\nu$, respectively. Applying the Fourier transform to both sides of Eq. \eqref{E:Boussinesq}, we have $\displaystyle{\tilde{p}_{\text{c}}(q) = \frac{1}{2} E^* q \tilde{h}(q)}$ \cite{Persson01, Persson08, Wang17}, where $p_{\text{c}}(x, y) \Leftrightarrow \tilde{p}_{\text{c}}(q)$. Let $S[p_{\text{c}}](q) = \displaystyle{\frac{(2 \pi)^2}{A_{\text{n}}}}|\tilde{p}_{\text{c}}(q)|^2$ be the PSD of $p_{\text{c}}(x, y)$, then we can find an important identity between $S[p_{\text{c}}](q)$ and $C(q)$, i.e., 
\begin{equation}\label{eq:identity_PSD}
S[p_{\text{c}}](q) = \frac{1}{4} (E^{*})^2 q^2 C(q).
\end{equation} 
Following Eq. \eqref{eq:variance_h_continuous}, the scale-dependent variance of $p_{\text{c}}(x, y)$, which is denoted by $V$, can be written as
\begin{equation}\label{eq:V_equ_1}
V = \displaystyle{2 \pi \int_{q_{\text{l}}}^{\zeta q_{\text{l}}} q S[p_{\text{c}}](q)} \text{d} q.
\end{equation}
Substituting Eq. \eqref{eq:identity_PSD} into Eq. \eqref{eq:V_equ_1}, we can find an alternative form for $V$ \cite{Manners06}, i.e.,
\begin{equation}\label{eq:V_equ}
V = \frac{\pi}{2} (E^*)^2 \int_{q_{\text{l}}}^{\zeta q_{\text{l}}} q^3 C(q) \text{d} q.
\end{equation}
Substituting Eq. \eqref{eq:V_gradient} into \eqref{eq:V_equ}, an important identity can be deduced \cite{Manners06, Yastrebov15}:
\begin{equation}\label{eq:VV_relation}
V = \frac{1}{4} (E^*)^2 \langle |\nabla h|^2 \rangle.
\end{equation}
Substituting Eq. \eqref{E:PSD} into Eq. \eqref{eq:V_gradient}, we can have the closed-form solution of $\langle (\nabla h)^2 \rangle$. Substituting it into the above identity, the closed-form solution of $V(\zeta)$ is obtained as
\begin{align}
V(\zeta) &=
\begin{cases}
\displaystyle{\frac{1}{8} \pi (E^*)^2 C_0 q_{\text{r}}^{-2H - 2} q_{\text{l}}^4 \left( \zeta^4 - 1 \right)} ~~~ \zeta q_{\text{l}} < q_{\text{r}}, \\
\displaystyle{\frac{1}{8} \pi (E^*)^2 C_0 \left[ q_{\text{r}}^{-2H - 2} \left( q_{\text{r}}^4 - q_{\text{l}}^4 \right) + \frac{2}{1 - H} \left( \zeta^{-2H + 2}q_{\text{l}}^{-2H + 2} - q_{\text{r}}^{-2H + 2} \right) \right]} ~~~ \zeta q_{\text{l}} \geq q_{\text{r}}.
\end{cases} \label{eq:Vp_closed_form}
\end{align}
The elastic strain energy, which is stored in the deformed body, at   complete contact is defined as
\[
U_{\text{el}} = \frac{1}{2} \int_{-L_x/2}^{L_x/2} \int_{-L_y/2}^{L_y/2} p_{\text{c}}(x, y) h(x, y) \text{d} x \text{d} y.
\]
Let the cross-correlation be
\[
\text{Corr}[p_{\text{c}}, h](x, y) = \frac{1}{A_{\text{n}}}\int_{-L_x/2}^{L_x/2} \int_{-L_y/2}^{L_y/2} p_{\text{c}}(x + x', y + y') h(x', y') \text{d} x' \text{d} y'.
\]
Hence, $U_{\text{el}} = \displaystyle{\frac{1}{2}} A_{\text{n}} \text{Corr}[p_{\text{c}}, h](0, 0)$. According to the correlation theorem, we have $\text{Corr}[p_{\text{c}}, h](x, y) \Leftrightarrow \displaystyle{\frac{(2 \pi)^2}{A_{\text{n}}} \tilde{p}_{\text{c}}(q_x, q_y) \tilde{h}^*(q_x, q_y)}$, where ``$\ast$" denotes the complex conjugate. Using the identity $\tilde{p}_{\text{c}}(q_x, q_y) = \displaystyle{\frac{1}{2}} E^* q \tilde{h}(q_x, q_y)$, as $L_x$ and $L_y$ are sufficiently large, $U_{\text{el}}$ per nominal contact area can be formulated as \cite{Persson08}
\begin{align}
U_{\text{el}} &= \frac{(2 \pi)^3}{E^*} \int_{q_{\text{l}}}^{\zeta q_{\text{l}}} |\tilde{p}_{\text{c}}(q)|^2 \text{d} q, \label{eq:Uel_complete_1} \\
&= \frac{\pi}{2} A_{\text{n}} E^* \int_{q_{\text{l}}}^{\zeta q_{\text{l}}} q^2 C(q) \text{d} q. \label{eq:Uel_complete_2}
\end{align}

\setcounter{figure}{0}
\setcounter{equation}{0}
\renewcommand{\thefigure}{C.\arabic{figure}}
\renewcommand{\theequation}{C.\arabic{equation}}
\renewcommand{\thetable}{C.\arabic{table}}
\section*{Appendix C. $B_n(\zeta) = \text{d}\langle p^n \rangle/\text{d}\zeta$, $n = 1, 2$}
Substituting Eq. \eqref{E:Bn_forward_KM_expansion} with $n = 1$ into Eq. \eqref{eq:mean_Bn} and using the change of variable: $\Delta p = p' - p$, we have
\begin{align}
B_1(\zeta) &= \lim_{\Delta \zeta \to 0} \frac{1}{\Delta \zeta} \int_0^{\infty} \int_0^{\infty} (p' - p) P_0(p', \zeta + \Delta \zeta)|p, \zeta) P_0(p, \zeta) \text{d} p' \text{d} p, \notag \\
&= \lim_{\Delta \zeta \to 0} \frac{1}{\Delta \zeta} \left[ \int_0^{\infty} p' P_0(p', \zeta + \Delta \zeta) \text{d} p' - \int_0^{\infty} p P_0(p, \zeta) A_{\text{t}}^*(p, \zeta, \Delta \zeta) \text{d}p \right]. \label{eq:C1}
\end{align} 
Due to the fact that $A_{\text{t}}^* \to 1$ once $\Delta \zeta \to 0$, Eq. \eqref{eq:C1} is written as
\begin{equation}
B_1(\zeta) = \lim_{\Delta \zeta \to 0} \frac{1}{\Delta \zeta} \left[ \langle p \rangle(\zeta + \Delta \zeta) - \langle p \rangle(\zeta) \right] = \frac{\text{d} \langle p \rangle}{\text{d} \zeta}.
\end{equation}
Similarly, 
\begin{align}
B_2(\zeta) &= \lim_{\Delta \zeta \to 0} \frac{1}{\Delta \zeta} \int_0^{\infty} \int_0^{\infty} (p' - p)^2 P_0(p', \zeta + \Delta \zeta)|p, \zeta) P_0(p, \zeta) \text{d} p' \text{d} p, \notag \\
&= \lim_{\Delta \zeta \to 0} \frac{1}{\Delta \zeta} \bigg\{ \int_0^{\infty} (p')^2 P_0(p', \zeta + \Delta \zeta) \text{d} p' - 2 \int_0^{\infty} p \left[ \int_0^{\infty} p' P_0(p', \zeta + \Delta \zeta|p, \zeta) \text{d} p' \right] P_0(p, \zeta) \text{d}p + \notag \\
& \int_0^{\infty} p^2 A_{\text{t}}^*(p, \zeta, \Delta \zeta) P_0(p, \zeta) \text{d}p \bigg\}.
\end{align}
Once $\Delta \zeta \to 0$, $P_0(p', \zeta  + \Delta \zeta|p, \zeta) \to \delta(p' - p)$. Therefore, $B_2(\zeta)$ can be further simplified to 
\begin{equation}
B_2(\zeta) = \lim_{\Delta \zeta \to 0} \frac{1}{\Delta \zeta} \left[ \langle p^2 \rangle(\zeta + \Delta \zeta) - \langle p^2 \rangle(\zeta) \right] = \frac{\text{d} \langle p^2 \rangle}{\text{d} \zeta}. 
\end{equation}

\end{document}